\documentclass[12pt]{article}
\usepackage{mathrsfs,amssymb,amstext,amsmath,amsthm,eucal}
\usepackage{graphicx,subfig}
\usepackage{longtable}
\usepackage[normalem]{ulem}
\usepackage{marvosym}
\usepackage{hyperref}
\usepackage{geometry}
\usepackage{authblk}
\usepackage[usenames,dvipsnames]{color}
\usepackage{enumerate} 

\newcommand{\grad}{\mathrm{d}}
\newcommand{\CC}{\mathbb{C}}
\newcommand{\RR}{\mathbb{R}}

\newcommand{\dvol}{\mathrm{dvol}} 
\newcommand{\lie}{\mathcal{L}}

\newcommand{\braket}[1]{\left\langle #1 \right\rangle}
\newcommand{\AdS}{\mathrm{AdS}}

\author{George Moutsopoulos\footnote{gmoutso@gmail.com}}
\affil{Department of Mathematics, Bogazici University}
\date{\today}
\title{Warped anti-de Sitter in 3d $(2,0)$ Supergravity}
\hypersetup{
  pdfkeywords={},
  pdfsubject={},
  pdfcreator={Emacs 24.3.1 (Org mode 8.2.7b)}}
\begin{document}

\maketitle

\begin{abstract}
We comment on the ubiquity of the so-called warped anti-de Sitter spacetimes in three-dimensional (2,0) supergravity theory. By using isometry-invariant tensors and simple counting, we prove their existence for arbitrary $(2,0)$ supergravity models suitably defined close to a minimal model. We also analyze their offshell supersymmetry and the supersymmetry of two geometric orbifolds.
\end{abstract}

\newpage
\tableofcontents
\section{Motivation}
In this note, we comment on the ubiquity of the so-called warped anti-de Sitter spacetimes in three-dimensional supergravity theory. These spacetimes are homogeneous vector-like deformations of anti-de Sitter, defined on the group manifold $\mathrm{SL}(2,\RR)$ such that the isometry algebra\footnote{These are the Killing vectors that leave invariant all gauge-invariant tensors of the solution, not only the metric.}\footnote{The subscript L (R) on the $\mathfrak{sl}(2,\RR)$ algebras denotes whether they generate the left (right) action on the group $\mathrm{SL}(2,\RR)$.} 
 of anti-de Sitter
\begin{equation}\label{eq:so22}
\mathfrak{so}(2,2)=\mathfrak{sl}(2,\RR)_L \oplus \mathfrak{sl}(2,\RR)_R
\end{equation}
is broken to its subalgebra ${ \mathfrak{iso}_k }$, which is the centralizer of a left-invariant vector $k \in \mathfrak{sl}(2,\RR)_R $ in $\mathfrak{so}(2,2)$,
\begin{equation}\label{eq:sl2plusR}
{ \mathfrak{iso}_k }=  \mathfrak{sl}(2,\RR)_L\oplus\RR\braket{k} 
=\left\{\xi\in \mathfrak{so}(2,2): [k,\xi]=0\right\}
~.
\end{equation}
The vector $k$ may in turn be timelike, spacelike or null with respect to the metric, which corresponds to an elliptic, hyperbolic or parabolic element. We accordingly call the warped anti-de Sitter spacetimes as spacelike, timelike or null warped anti-de Sitter. These solutions have appeared as supersymmetric solutions of the $(1,1)$ supergravity model in \cite{deger_supersymmetric_2013} and recently as supersymmetric solutions of the $(2,0)$ model in \cite{prep}. The motivating question of this note is  
``\textit{how natural is the existence of these solutions?}''. 
By using isometry-invariant tensors and simple counting, we will prove their existence for arbitrary $(2,0)$ supergravity models suitably defined close to a minimal model.

Although one may argue about the existence of these solutions for more general theories, we will concentrate firmly on $(2,0)$ anti-de Sitter supergravity. The supersymmetry multiplet consists of a dreibein, a complex gravitino, two gauge fields and a real scalar. Their supersymmetry variations close off-shell, so one can make a distinction between supersymmetry results that hold for a solution to an arbitrary $(2,0)$ supergravity model and the solution of a specific  $(2,0)$ supergravity model. That is, we make here the important distinction between an (offshell) \textit{background} and an (onshell) \textit{solution} of a particular model. We will use a minimal model that contains the Einstein-Hilbert term, a cosmological constant, and a gravitational topological term. This model has been called the $(2,0)$ cosmological topologically massive supergravity \cite{prep}, where the term massive refers to a massive gravity mode \cite{DeserTMG} and cosmological due to the cosmological constant~\cite{gibbons_general_2008}. The minimal model provides us with a rich set of warped anti-de Sitter solutions. 

There are three instances where warped anti-de Sitter has appeared with regard to $(2,0)$ supergravity
\begin{enumerate}[a)]
\item A half-supersymmetric solution is found in \cite{prep} that corresponds to timelike warped anti-de Sitter.
\item All maximally supersymmetric backgrounds are found in \cite{knodel_rigid_2015}, which include timelike, spacelike and null warped anti-de Sitter.
\item The purely gravitational solutions that are discussed for instance in \cite{anninos_warped_2009,chow_classification_2010}, which according to the results of  \cite{gibbons_general_2008} are not supersymmetric.
\end{enumerate}
We make the following observations about these backgrounds and solutions. The half-supersymmetric solution in \cite{prep} is produced from what is called there a general constancy Ansatz that is imposed on supersymmetric solutions of the minimal model. It is generic in the sense that a ratio of coupling constants need only satisfy a certain inequality. The maximally supersymmetric spacelike, timelike and null warped anti-de Sitter of \cite{knodel_rigid_2015} was also produced in \cite{prep} from what was called there the special constancy Ansatz, again imposed on supersymmetric solutions. The special constancy Ansatz produces this solution only if a ratio of coupling constants of the minimal model is at a fixed value. These observations motivate a couple of questions with regard to the landscape of warped anti-de Sitter as supersymmetric backgrounds and as solutions to $(2,0)$ supergravity models.

Firstly, we ask whether ``there are any other warped anti-de Sitter solutions to be found in the minimal model''. Indeed, the solutions of \cite{prep} are the result of an Ansatz imposed on supersymmetric solutions, the constancy Ansatz, so there may be other warped anti-de Sitter geometries that evade the method of \cite{prep}, which may or may not be supersymmetric. 
We prove that the supersymmetric  warped anti-de Sitter solutions in \cite{prep} contains all supersymmetric warped anti-de Sitter solutions of the minimal model, and there is only one more class of  non-supersymmetric warped anti-de Sitter solutions, the purely gravitational solutions of case (c) above.

\begin{figure}
   \centering
   {\includegraphics{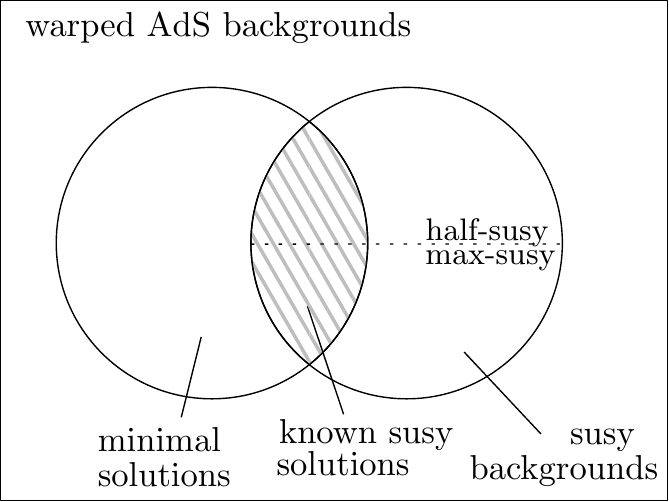}}
  \caption[wAdS solutions vs susy]{In the space of warped anti-de Sitter backgrounds, two sets are sketched: all such geometries that are solutions to the minimal model (left circle), all such geometries that preserve some supersymmetry (right circle). 
We show that all minimal solutions, both supersymmetric and non-supersymmetric, are known. We also find the precise amount of supersymmetry that supersymmetric backgrounds admit. Up to homotheties, the complement of the right circle is in fact empty.}\label{fig:venn}
\end{figure}

Perhaps more interesting than our result is the method we use. The method we employ to find all warped anti-de Sitter solutions of the minimal model draws\footnote{That work dealt with the non-relativistic holography of $1/N$ corrections.}
 inspiration from \cite{Adams:2008zk}. It allows us to answer the motivating question ``how natural these solutions are'' not only for the minimal model but for other $(2,0)$ models or even more general theories. Specifically, we call a warped anti-de Sitter background a geometry on the group manifold $\mathrm{SL}(2,\RR)$ where all fields are invariant under the isometry $\mathfrak{iso}_k$ that was described in \eqref{eq:sl2plusR}. We may then decompose all equations of motion into its isometry-invariant components, which comprise a finite dimensional space. This reduces the problem to a non-linear analysis of five algebraic equations for five algebraic constants. This provides a definite answer for both the minimal model and more general theories parametrically close to the minimal model. 

The second question we answer is ``which of the backgrounds preserve supersymmetry and how much''. We thus find conditions for allowing none, one, or two complex supersymmetries. With regard to maximally supersymmetric backgrounds we are confirming parts of the results of \cite{knodel_rigid_2015} and with regard to the half-supersymmetric solution of the minimal model we are confirming a part of the results of \cite{prep}. The essential novelty of this section is then the conditions for allowing half-supersymmetric backgrounds, which may or not be solutions to any given theory. In any case, we provide here a supersymmetry analysis, namely the integrability of the Killing spinor equations, by using invariant objects of the geometry. We conclude with the supersymmetry preserved by some isometric quotients. 

A schematic of the two questions and their answer is given in figure \ref{fig:venn}. In sections \S\ref{sec:preliminaries} and \S\ref{sec:wads} we respectively introduce $(2,0)$ supergravity and define appropriately the warped anti-de Sitter backgrounds. In section \S\ref{sec:assolutions} we answer the question of how natural these backgrounds are as solutions to $(2,0)$ supergravity models and in section \S\ref{sec:assusy} we analyze their supersymmetry. Finally, in section 
\S\ref{sec:asquotients}
we conclude with a discussion on the Killing superlagebras and their supersymmetric quotients. For ease of reading, we will relegate details of the null case to the appendix. We will not use explicit coordinates to parametrize $\mathrm{SL}(2,\RR)$, coordinate forms of the Maurer-Cartan one-forms and of the metrics can be found in \cite{jugeau_accelerating_2011}, see also \cite{prep}.

\section{(2,0) Supergravity}
\label{sec:preliminaries}
The offshell supersymmetry transformations of the (2,0) supergravity multiplet were given in \cite{kuzenko_three-dimensional_2011,kuzenko_three-dimensional_2014,alkac_massive_2015} by fixing the scale of a three-dimensional superconformal multiplet. The supergravity multiplet then consists of a dreibein $e^a_{\mu}$, a complex gravitino $\psi_\mu$ that transforms under a local $U(1)_R$ with gauge field $V_{\mu}$, an abelian gauge field $C_{\mu}$ and a real scalar $D$. We define the field strengths $G=\grad C$ and $F=\grad V$. The mutiplet transform under a complex Grassmann-odd spinor $\epsilon$ as
\begin{align}
\delta e_{\mu}^{a}& =  \frac{1}{2}i \,\bar{\epsilon}\,\gamma^{a}\,\psi_{\mu}+\text{h.c.}
 \\
\delta\psi_{\mu}& =  \hat{D}_{\mu}\epsilon
\\
\delta C_{\mu}& =  \frac{1}{4}  \,\bar{\epsilon}\,\psi_{\mu}+ \text{h.c.}
\\
\delta V_{\mu}& =   \bar{\epsilon}\,\gamma^{\nu}\hat{D}_{[\mu}\psi_{\nu]} -\frac{1}{4}  \bar{\epsilon}\,\gamma_{\mu}\gamma^{\nu\rho}\hat{D}_{\nu}\psi_{\rho}
- i \bar{\epsilon}\,\hat{G}\,\psi_\mu - D\bar\epsilon\, \psi_\mu +\text{h.c.}
 \\
\delta D & =  -\frac{1}{8} i\bar{\epsilon}\,\gamma^{\mu\nu}\hat{D}_{\mu}\psi_{\nu} +\text{h.c.}
\end{align}
where the supercovariant derivative is
\begin{equation}
\hat{D}_{\mu}\epsilon := \left(
\nabla_{\mu}
-i V_{\mu}- i \, \gamma_{\mu} \hat{G}
-\gamma_{\mu}\,D\right)\epsilon~ \label{eq:D}
\end{equation}
and $\hat{G}$ is the supercovariant field strength of $C$. 

Our conventions for the gamma matrices are $\{\gamma_a,\gamma_b\}=+2\eta_{ab}$ for a mostly plus signature metric and $\gamma_{\underline{0}\underline{1}\underline{2}}=\epsilon_{\underline{0}\underline{1}\underline{2}}=\zeta$ in an orthonormal basis with indices  $\underline{a},\underline{b}=\underline{0},\underline{1},\underline{2}$ is a sign $\zeta=\pm1$ that depends on the volume form
\begin{equation}
\dvol = \zeta \, \theta^{\underline{0}}\wedge \theta^{\underline{1}}\wedge \theta^{\underline{2}}~.
\end{equation} Then, $p$-forms act on spinors via their image in the Clifford module, e.g. $G\epsilon=\frac{1}{2}G_{ab}\gamma^{ab}\epsilon$, $k\epsilon=k_{\mu}\gamma^{\mu}\epsilon$ and $\dvol \, \epsilon=-\epsilon$. With these conventions, we note the useful relation 
\begin{equation}
k\epsilon =-\left(\ast k \right)\epsilon
\label{eq:hodgegamma}
\end{equation}for a one-form $k$ and that $\ast^2=-1$ on the whole of the exterior algebra. We will  identify\footnote{The identification is compatible with the action of the Levi-Civita derivative along any vector $X$ and the action of the Lie derivative along Killing vectors, e.g. $\nabla_X k$ and $g(\nabla_X k,-)$ are unambiguous.} a vector $k$ with its metric-dual one-form $g(k,-)$, so that an expression such as $\grad k$ also makes sense.  The spinor inner product in the supersymmetry transformations is anti-hermitian for commuting spinors, but spinor bilinears will not enter our analysis. 

The $(2,0)$ multiplet transforms under a local $e^{i\phi}\in U(1)_R$ as
\begin{align}
\psi_{\mu} &\mapsto e^{i\phi} \psi_{\mu}\\
V_{\mu} &\mapsto V_{\mu} + \partial_{\mu} \phi \label{eq:Vtrans}
\end{align}
The supersymmetry parameter $\epsilon$ transforms with the same weight as $\psi_{\mu}$,
\begin{equation}
\epsilon \mapsto  e^{i\phi} \epsilon~, \label{eq:KStransU1}
\end{equation} and it follows that the supercovariant derivative $\hat{D}_{\mu}$ in \eqref{eq:D} is $U(1)_R$ covariant. We will make the distinction between a (bosonic) background that is given by the data $(g,V,G,D)$ that might or not be a solution to an actual model, and a (bosonic) solution to a given  $(2,0)$ supersymmetric model that is a background satisfying the model's equations of motion  with $\psi_{\mu}\equiv0$. A Killing spinor of a background  $(g,V,G,D)$ is such that
\begin{equation}
{D}_{\mu}\epsilon :=  \left(
\nabla_{\mu}
-i V_{\mu}- i \, \gamma_{\mu} {G}
-\gamma_{\mu}\,D\right)\epsilon= 0~. \label{eq:KS}
\end{equation}
Clearly, if $\epsilon$ is a Killing spinor of a background then so is $i\epsilon$ and Killing spinors form a finite-dimensional complex space of maximal complex dimension two. Therefore, the supersymmetry preserved by the background is either none, half or full. Note that the distinction between a background and a solution is not meant to imply that any background is a solution to some suitable $(2,0)$ model. However, we may tacitly consider Killing spinors as the fermionic counterpart of Killing vectors and the supersymmetry transformations as the counterpart of infinitesimal diffeomorphisms, without a reference to any given theory.

We will eventually want to connect to solutions of specific models. We choose the topological massive $(2,0)$ theory that is minimal in the sense that it provides a rich set of warped anti-de Sitter solutions. The bosonic section of the lagrangian 
 is given by
\begin{equation}
\begin{aligned}
L &= M \left( 2 D - \epsilon^{abc}C_a G_{bc} \right) \\
&+ R + 4 G_{ab}G^{ab} - 8 D^2 -8 \epsilon^{abc}C_a \partial_b V_c \\
& - \frac{1}{4\mu} \epsilon^{\mu\nu\rho}\left( R_{\mu\nu}{}^{ab} \omega_{\rho ab} +\frac{2}{3} \omega_\mu{}^{ab}\omega_{\nu b}{}^c\omega_{\rho ca} - 8 V_\mu \partial_\nu V_\rho \right)~,\label{eq:lagrangian}
\end{aligned}
\end{equation}
and contains all  $(2,0)$ supersymmetry invariant terms up to third-order derivatives \cite{alkac_massive_2015}.  The lagrangian depends on two real parameters, $\mu$ and $M$, which we may vary freely in order to fix the model in the theories' parameter space. We will ultimately consider models that are parametrically close to the minimal model, in the sense of adding higher-order terms with small enough coefficients or by varying $\mu$ and $M$ from some fixed values.

The equations of motion from the lagrangian \eqref{eq:lagrangian} are
\begin{align}
M&=8D \label{eq:MD}\\
- \grad \ast G &= \frac{1}{2}F + \frac{M}{4} G \label{eq:vector1}\\
F &= 2 \mu \, G \label{eq:vector2}
\end{align}
and the Einstein equation
\begin{equation}
R_{\mu\nu} - \frac{1}{2} \left(R+ 2 MD - 8 D^2 \right) g_{\mu\nu} + \frac{1}{\mu} C_{\mu\nu} + 8 G_{\mu a} G_{\nu}{}^a  - 2 G_{ab}G^{ab} g_{\mu\nu}=0~, \label{eq:Einstein}
\end{equation}
where we define the symmetric and traceless  Cotton tensor
\begin{equation}
C_{\mu\nu} = - \epsilon_\mu{}^{ab}\left( \nabla_bR_{\nu a}-\frac{1}{4}g_{\nu a} \partial_b R\right)~.\label{eq:cotton}
\end{equation}
Note that $D$ and $F$ are fixed algebraically by \eqref{eq:MD} and \eqref{eq:vector1} and can be replaced back into the lagrangian. In this model the two $U(1)$ bundles are identified in \eqref{eq:vector2} and in fact one might locally integrate the gauge field equations of motion into a single massive vector equation. 
A more important note to us is that, a minimal model solution depends on the choice  of orientation, e.g. its sign appears in the definition of the Cotton tensor \eqref{eq:cotton} and the Hodge operation in \eqref{eq:vector1}. It is also useful to recall that, if we set the gauge fields to zero, the model and supersymmetry transformations coincide with the purely gravitational $\mathcal{N}=1$ cosmological topologically massive supergravity, whose supersymmetric solutions where solved for in \cite{gibbons_general_2008}.

\section{As Backgrounds}
\label{sec:wads}
We will now introduce the warped anti-de Sitter backgrounds. Let us momentarily note that there are various\footnote{See \cite[\S4.1]{moutsopoulos_homogeneous_2013} and references therein for a classification.} types of $\mathrm{SL}(2,\RR)$-invariant metrics, more general to warped anti-de Sitter. However, warped anti-de Sitter is an $\mathrm{SL}(2,\RR)$-invariant deformation of anti-de Sitter that may be characterized as a vector-like deformation, it enjoys special properties such as the supersymmetry in \cite{deger_supersymmetric_2013,prep,knodel_rigid_2015}, and in fact appears frequently in the literature, see e.g. \cite{chow_classification_2010}. 

We will use the ``lorentzian'' conventions of \cite{jugeau_accelerating_2011} for the $\mathfrak{sl}(2,\RR)$ bases. 
On $\mathrm{SL}(2,\RR)$ we define the left-invariant Mauer-Cartan basis of one-forms $\tau^a$ and their dual left-invariant vector fields $L_a$, $\tau^a(L_b)=\delta^a_b$, which generate the right action $\mathrm{SL}(2,\RR)_R$ on the group manifold. They satisfy
\begin{equation}
[L_a,L_b]= f_{ab}{}^cL_c \Longleftrightarrow \grad \tau^a+ \frac{1}{2}f_{bc}{}^a \tau^b \wedge \tau^c =0~,
\end{equation}
where we use the convention that $f_{012}=1$ is completely antisymmetric and the $0,1,2$ labels on $L_a$, $\tau_a$ and $f_{abc}$ are raised and lowered with a mostly plus Minkowski metric $\eta_{ab}$. We also define the basis of right-invariant vector fields $R_a$ such that
\begin{equation}
[R_a,R_b]= f_{ab}{}^cR_c~,
\end{equation}
which commute with the $L_a$ and generate the left action $\mathrm{SL}(2,\RR)_L$, whence
\begin{equation}\label{eq:Rtau}
\lie_{R_a}\tau^b = 0~.
\end{equation} The three-dimensional anti-de Sitter metric is the Cartan-Killing metric on the group manifold $\mathrm{SL}(2,\RR)$
\begin{equation}\label{eq:ads}
g_{\AdS} = \frac{\ell^2}{4} \left( -  \tau_0^2 + \tau_1^2 + \tau_2^2 \right) 
\end{equation}
and $\ell$ is the cosmological radius. Upon using that the $\tau_a$ transform in the $\RR^{1,2}$ representation of $\mathfrak{sl}(2,\RR)_R=\mathfrak{so}(1,2)$ as an orthonormal frame,
\begin{equation}\label{eq:Ltau}
\lie_{L_a} \tau^b = - f_{ac}{}^b \tau^c~,
\end{equation}
we may confirm that the anti-de Sitter metric is indeed invariant under both the left \eqref{eq:Rtau} and right \eqref{eq:Ltau} action, as stated in \eqref{eq:so22}.

Spacelike, timelike and null warped anti-de Sitter is refered to the deformation of the anti-de Sitter metric by a constant times $k^{\sharp} \otimes k^{\sharp}$, where $k^{\sharp}$ is a left-invariant one-form,
\begin{equation}
g = g_{\AdS} + \text{const.} \times  k^{\sharp} \otimes k^{\sharp}~.
\end{equation} We may use the $\mathfrak{sl}(2,\RR)_R$ action on $k^{\sharp}$ to bring it to the form
\begin{equation}
k^{\sharp} = \text{const'.} \times \begin{cases} \tau_0 & \text{timelike}
\\
\tau_1 & \text{spacelike} \\
\tau_0 + \tau_1 &  \text{null}\end{cases}~,
\end{equation}
where the flip $\tau_0 - \tau_1 \mapsto \tau_0 + \tau_1$ is also an inner automorphism of the algebra (by also exchanging the sign of $\tau_2$) and $\tau_0+\tau_1$ may be rescaled freely by a lorentzian boost in $\mathfrak{sl}(2,\RR)_R$, 
\begin{equation}
e^{\zeta \lie_{L_2}}\left(\tau_0 + \tau_1\right)=e^{-\zeta}\left(\tau_0 + \tau_1\right)~.
\end{equation}
We then have the following cases of vector-like homogeneous deformations of anti-de Sitter, the so-called warped anti-de Sitter metrics
\begin{equation}\label{eq:wads}
g =  \begin{cases}  \frac{\ell^2}{4} \left( - \lambda^2\tau_0^2 + \tau_1^2 + \tau_2^2 \right) & \text{timelike}
\\
\frac{\ell^2}{4} \left( - \tau_0^2 + \tau_1^2 + \lambda^2 \tau_2^2 \right) & \text{spacelike} \\
g_{\AdS} \pm \frac{\ell^2}{4}\left(\tau_0 + \tau_1 \right)^2 &  \text{null}\end{cases}~,
\end{equation}
where $\lambda$ and $\ell$ are constants. Note that by null warped anti-de Sitter we call both a negative and a positive sign deformation.

In each of the three cases, we may find the left-invariant vector $k$ that leaves $k^{\sharp}$ invariant, i.e. $\lie_k k^{\sharp}=0$, and thus leave the metric $g$ invariant. 
In fact, we may unambiguously relate $k^{\sharp}$ and $k$ via the metric $k^{\sharp}=g(k,-)$. Let us use this convention for any vector $X$ and identify it with its metric dual one-form $X\stackrel{=}{\mapsto} g(X,-)$. We henceforth normalize $k$ to be 
unit $\pm1$ in the spacelike / timelike case and  $k =\ell/2\,( \tau_0 + \tau_1 )$ in the null case,
\begin{equation}
k = \begin{cases} \frac{2}{\ell \lambda} L_0 & \text{timelike}
 \\
 \frac{2}{\ell \lambda} L_2 & \text{spacelike} \\
\frac2{\ell}\left(L_0 + L_1\right) &  \text{null}\end{cases}~.
 \end{equation}
The isometry of the warped anti-de Sitter metrics is the centralizer $\mathfrak{sl}(2,\RR)_L\oplus \RR\braket{k}$ of $k$ in $\mathfrak{so}(2,2)$, as first mentioned in \eqref{eq:sl2plusR}.

We will call a warped anti-de Sitter background a background where all fields are invariant under the isometries of the warped anti-de Sitter metric and not only the metric. 
Let us explain what that means in the case of the spacelike warped anti-de Sitter. Let us denote the tangent bundle $TM$ with a slight abuse of notation as the space of vector fields, i.e. sections of the former. We say that a vector field $X\in TM$ is isometry invariant, and write
\begin{equation}
X \in \left( TM \right)^{\text{Iso}}~,
\end{equation}
if it is left invariant by all isometries, i.e. under the infinitesimal action it satisfies
\begin{equation}
\lie_{\xi} X = 0 \text{ for all }\xi\in \mathfrak{sl}(2,\RR){\oplus}\RR~.
\end{equation}
Let 
us expand the vector field $X$ in the left-invariant Maurer-Cartan basis $\tau_a$ (more precisely expand the metric dual one-form of $X$)
\begin{equation}
X = a\, \tau_0 + b\, \tau_1 + c\, \tau_2 \in  \left( TM \right)^{\text{Iso}} ~,
\end{equation}
where $a$, $b$ and $c$ are generically functions on $\mathrm{SL}(2,\RR)$. Since $\mathfrak{sl}(2,\RR)_L\subset{ \mathfrak{iso}_k }$ leaves the MC forms invariant, i.e.
 \begin{equation}
 \lie_{\xi}X = (\lie_{\xi}a)\, \tau_0 + (\lie_{\xi} b)\, \tau_1 + (\lie_{\xi} c)\, \tau_2 = 0\quad \text{for all  }\xi\in\mathfrak{sl}(2,\RR)~,
\end{equation} 
and acts freely on $\mathrm{SL}(2,\RR)$, we deduce that the $a$, $b$ and $c$ are constants. But the remaining Killing vector $k$ of spacelike warped anti-de Sitter rotates the $\tau_0$ and $\tau_1$ in the vector representation $\RR^{1,1}$ of $\mathfrak{so}(1,1)$, hence we must have $b=a=0$. That is, the space of isometry-invariant vector fields is a finite-dimensional vector space that is spanned by $\tau_2$,
\begin{equation}
 \left( TM \right)^{\text{Iso}} = \RR = \left\{ a\, \tau_2, \,\, a\in\RR\right\}~.\label{eq:iso1}
\end{equation}
We will thus require the gauge field $V$ to be proportional\footnote{Since $V$ is a gauge field we require a gauge choice such that this is true.} to $\tau_2$ up to a constant. One can similarly prove that the space of isometry-invariant two-forms is one-dimensional, 
\begin{equation}
\left( \Lambda^2 M \right)^{\text{Iso}} =  \RR = \left\{ a\, \tau_0 \wedge \tau_1, \,\, a\in\RR\right\}~,
\end{equation}
and the space of isometry-invariant symmetric tensors is two-dimensional, 
 \begin{equation}\label{eq:s2inv}
  \left( S^2 M \right)^{\text{Iso}} = \RR^2 = \left\{ a\, \tau_2^2 + b\, \left(-\tau_0^2+\tau_1^2\right), \,\,a,b\in\RR\right\}~.
 \end{equation}
Analogous results exist for timelike warped anti-de Sitter, e.g. in that case we will take $V$ proportional to $\tau_0$ up to a constant, and similarly for null warped anti-de Sitter.

The unique geometry for the fields on the manifold $\mathrm{SL}(2,\RR)$ such that the isometry group of spacelike warped anti-de Sitter leaves them invariant, is parametrized by the spacelike warped anti-de Sitter background
\begin{align}\label{eq:wadss}
g &= \frac{{ \ell }^2}{4} \left( - \tau_0^2 + \tau_1^2 + \lambda^2 \tau_2^2 \right) &
V &= v \, \theta^{\underline{2}} \\
G &= { g_1 } \, \theta^{\underline{0}} \wedge \theta^{\underline{1}} 
&
D &= d~,
\end{align}
where ${ \ell }>0$, $\lambda>0$, $v$, ${ g_1 }$ and $d$ are constants, 
the orthonormal frame we choose is
\begin{equation}
\theta^{\underline{a}} = \frac{{ \ell }}{2} \times \begin{cases} \tau_0 & \underline{a} = \underline{0}
\\
\tau_1
&\underline{a}=\underline{1} 
\\
\lambda \tau_2
&\underline{a}=\underline{2} \end{cases}~,
\end{equation}
and we define the sign $\epsilon_{\underline{0}\underline{1}\underline{2}}=\zeta$.

Similarly, the timelike warped anti-de Sitter background is
\begin{align}\label{eq:wadst}
g &= \frac{{ \ell }^2}{4} \left( - \lambda^2\tau_0^2 + \tau_1^2 + \tau_2^2 \right) &
V &= v \, \theta^{\underline{0}} \\
G &= { g_1 } \, \theta^{\underline{1}} \wedge \theta^{\underline{2}}
&
D &= d~,
\end{align}
where  ${ \ell }>0$, $\lambda>0$, $v$, ${ g_1 }$ and $d$ are again constants, 
the orthonormal frame we choose is
\begin{equation}
\theta^{\underline{a}} = \frac{{ \ell }}{2} \times \begin{cases} \lambda\tau_0 & \underline{a} = \underline{0}
\\
\tau_1
&\underline{a}=\underline{1} 
\\
 \tau_2
&\underline{a}=\underline{2} \end{cases}~,
\label{eq:lasttimeAn}
\end{equation}
and we define the sign $\epsilon_{\underline{0}\underline{1}\underline{2}}=\zeta$. This is again the unique geometry for the  fields that the isometry group of timelike warped anti-de Sitter leaves invariant. 

Finally, we present the null warped anti-de Sitter background: 
\begin{align}\label{eq:metricnulltwo}
g &= \frac{{ \ell }^2}{4} \left( - \tau_0^2 + \tau_1^2 + \tau_2^2  \pm \left( \tau_0 + \tau_1\right)^2 \right) &
V &= v \frac{\ell}{2} \left( \tau_0 + \tau_1\right) \\
G &= { g_1 } \frac{\ell}{2}\,  \grad \left( \tau_0 + \tau_1\right)
&
D &= d~,
\end{align}
where  ${ \ell }>0$, $v$, ${ g_1 }$ and $d$ are again constants, 
and we define the sign $\zeta$ via the volume form 
\begin{equation}
\dvol= \zeta \frac{\ell^3}{8} \tau_0\wedge\tau_1\wedge\tau_2~.
\end{equation}
Note that we are presenting here two different null vector deformations that are differentiated by the sign $\pm1$.

\section{As Solutions}
\label{sec:assolutions}
The details of null warped anti-de Sitter is given in the appendix \ref{sec:null}. For reasons of brevity, we will expose both spacelike and timelike warped anti-de Sitter at the same time and differentiate\footnote{Note that the two $k$'s generate elliptic (compact) and hyperbolic (non-compact) subgroups of $\mathrm{SL}(2,\RR)$ respectively. However, at the level of Lie algebras where we work, $k$ may smoothly vary from being spacelike to being null and then timelike.} the two with an upper or lower sign respectively. 
The two Ans\"atze are parametrized by the five parameters $(\lambda,{ \ell },{ g_1 },v,d)$ and two signs, see \eqref{eq:wadss}-\eqref{eq:lasttimeAn}. We will first find solutions in the minimal model and then generalize to other theories.

Any tensor made out of the fields of the backgrounds is necessarily invariant under the isometries. Consider for instance the Ricci tensor and the Cotton tensor, which are made out of the metric, two symmetric tensors (the Cotton tensor is also traceless). The space of isometry-invariant symmetric tensors is a finite-dimensional vector space
 that is spanned by two elements: the metric $g$ and $k \otimes k$, see for instance \eqref{eq:s2inv}.  Accordingly\footnote{Such relations were also given in \cite{chow_classification_2010}. These equations were produced here with the help of a computer.}, the Ricci and Cotton tensors decompose in this basis as
\begin{align}
\text{Ricci} &= \frac{2}{{ \ell }^2} \left( \lambda^2 - 2 \right) g \pm \frac{4}{{ \ell }^2}\left(1-\lambda^2\right) k \otimes k \\
\text{Cotton} &=- \zeta\frac{4 \lambda}{{ \ell }^3} (\lambda^2-1) \, g \pm \zeta \frac{12 \lambda}{{ \ell }^3} \left( \lambda^2 -1 \right) k \otimes k~.
\end{align}
The energy-momentum tensor in the Einstein equation that is proportional to $G^2$ also decomposes into two terms proportional to $g$ and $k \otimes k$ with constant coefficients. That is, the entire Einstein equation has only two non-trivial components, respectively those of $g$ and $k \otimes k$, with constant coefficient that we set to zero
\begin{align}
\lambda^2 - 4 D^2 { \ell }^2- \frac{4 \lambda \zeta}{\mu { \ell }}(\lambda^2-1) \mp 
4 { \ell }^2 { g_1 }^2 &=0 \label{eq:alg1}\\
\pm \frac{4}{{ \ell }^2} (1-\lambda^2) \pm \frac{12\lambda}{\mu { \ell }^3}(\lambda^2-1)\zeta +8 { g_1 }^2 &=0~.\label{eq:alg2}
\end{align}
The two vector field equations for $V$ and $G$ also decompose into the space of isometry-invariant one-forms, which is a one-dimensional space that is spanned by $k$ alone, see for instance \eqref{eq:iso1}. We thus get two more algebraic equations from the vector equations. The equation for $F$ fixes $v$ in terms of ${ g_1 }$, 
\begin{equation}
\pm \frac{2\lambda}{{ \ell }} v = 2 \mu \, { g_1 }~,\label{eq:alg3}
\end{equation}
whereas the equation for $G$ gives 
\begin{equation}
\left(\mu+2d\right) g_1 = \frac{2\lambda}{{ \ell }}\zeta \, g_1~.\label{eq:alg4}
\end{equation}
The equation of motion for $D$ fixes 
\begin{equation}
d=\frac{M}{8}~.\label{eq:alg5}
\end{equation}
We thus have five algebraic equations, \eqref{eq:alg1}-\eqref{eq:alg5}, we can use to solve for five unknown parameters $(\lambda,{ \ell },{ g_1 },d,v)$.

Note that only \eqref{eq:alg3} involves $v$, whence we can solve for $v$ easily, and from \eqref{eq:alg5} we solve for $d$. The algebraic equations have four types of solution for the remaining parameters:
\begin{enumerate}[(I)]
\item \label{item:AdS} anti-de Sitter, for which
\begin{equation}
g_1=0~, \quad \lambda=1 \quad\text{and} \quad { \ell } = \frac{1}{8M}
\end{equation}
Note how anti-de Sitter requires all gauge fields turned off so that its enhanced isometry $\mathfrak{so}(2,2)$ is unbroken.
\item \label{item:general}Generic $(M,\mu)$ gives a real-parameter solution only for the timelike case
\begin{align}
\lambda &= \frac{(M+4\mu)}{\sqrt{M(M-4\mu)}}\zeta\\
{ \ell } &= \frac{8}{\sqrt{M(M-4\mu)}} \\
|{ g_1 }| &= \frac{ \left|3M + 4\mu\right|}{16} ~.
\end{align}
Note that it necessitates 
\begin{equation}
M(M-4\mu)>0~.\label{eq:inequality}
\end{equation}
The parameter $\lambda$ can be both squashed $\lambda<1$ or stretched $\lambda>1$.
\item \label{item:special}For $M=4\mu$ both timelike and spacelike warped anti de Sitter can be found. It can be parametrized by
\begin{align}
{ \ell }&= 4 \frac{\lambda}{M}\zeta \\
{ g_1 }^2 &= \mp\frac{\lambda^2-1}{16 { \lambda }^2}M^2~,
\end{align}
for $\lambda$ a free parameter. Note how this requires $\lambda<1$ ($\lambda>1$) for spacelike (timelike) warped anti-de Sitter. 
\item \label{item:pureTMG}The purely gravitational timelike and spacelike warped anti-de Sitter, with
\begin{align}
\lambda &= \frac{8 \mu}{\sqrt{27 M^2 + 16\mu^2}} \zeta\\
{ \ell } &= \frac{24}{\sqrt{27 M^2 +16 \mu^2}} \\
g_1 &= 0~.
\end{align}
All vector fields are turned off.
\end{enumerate}
We henceforth stop grouping the spacelike and timelike warped anti-de Sitter solutions with a sign. 

By using isometry invariant tensors, the equations of motion decompose into five algebraic equations for five unknown parameters that yield four types of solutions. Rather disappointingly, they comprise all known spacelike and timelike warped anti-de Sitter solutions \cite{prep,chow_classification_2010}. Anti-de Sitter (type \ref{item:AdS}) is maximally supersymmetric. For generic $(\mu,M)$ there is one purely gravitational spacelike warped anti-de Sitter (type \ref{item:pureTMG}) and one purely  gravitational timelike warped anti-de Sitter (type \ref{item:pureTMG}), and if \eqref{eq:inequality} holds then one timelike warped anti-de Sitter solution (type \ref{item:general}) that is according to \cite{prep} half-supersymmetric. From the results of \cite{gibbons_general_2008} about $\mathcal{N}=1$ purely gravitational TMG, both type \ref{item:pureTMG} solutions are not supersymmetric. 
For $M=4\mu$ there is a line of solutions (type \ref{item:special}), which according to \cite{prep} coincides with the maximally supersymmetric backgrounds of \cite{knodel_rigid_2015}. At $3M+4\mu=0$ the generic solution type \ref{item:general} and the purely gravitational solution \ref{item:pureTMG} coincide with anti-de Sitter.

The null warped anti-de Sitter case is treated in the appendix \ref{sec:null} and we summarize it here. Both negative and positive deformed null warped anti-de Sitter minimal model solutions are allowed for $4|\mu|=3|M|$, and only the negative sign deformation for $\mu=4M$. The  $4|\mu|=3|M|$ solutions are purely gravitational. Besides these, there are no other solutions. It should be possible to see some of the null warped anti-de Sitter solutions as limits of the spacelike and timelike warped anti-de Sitter solutions, type \ref{item:general} and \ref{item:special}, see \cite{jugeau_accelerating_2011}. For instance, the limit for type  \ref{item:special} is manifest in \cite{prep}, with $c\rightarrow 2\mu$ there.
 
We will now show that the method of decomposing into isometry invariants will yield type \ref{item:general} and \ref{item:pureTMG} solutions for any putative $(2,0)$ theory that is suitably close to the minimal model. First notice that by including higher order terms, the equations of motion still decompose into five algebraic equations for five unknown parameters  $({ \ell },\lambda,{ g_1 },v,d)$
\begin{equation}
F_i({ \ell },\lambda,{ g_1 },v,d; \mu_I)=0~,\quad i=1,\dots,5~.
\end{equation}
The $(\mu_I) = ( M, \mu, \dots, \mu_N)$ parametrize an arbitrary number $N$ of higher-order terms of the lagrangian. In order to argue for the existence of a solution to these equations, we will consider a finite deformation from the minimal model. Suitably close has the meaning of adding small enough higher order corrections to the lagrangian \eqref{eq:lagrangian} and these terms will only shift the parameters of the minimal model solutions.  This perspective is influenced by~\cite{Adams:2008zk}.

Consider first linearizing to first order the five equations with higher order terms with a derivation $\delta$ and evaluated at our theory parameters $(\mu,M)$ and solution the parameters $({ \ell },\lambda,{ g_1 },v,d)$ that we found earlier, e.g. for the type \ref{item:general} solution. We derive this way a $5\times5$ linear system for $(\delta \lambda_i) = (\delta { \ell }, \delta \lambda, \delta { g_1 }, \delta v, \delta d)$ in terms of an arbitrary number of linearized theory parameters $(\delta\mu_I) = (\delta M$, $\delta \mu$, $\dots, \delta\mu_N)$:
\begin{equation}\label{eq:linmat}
\sum_{j=1}^5\frac{\partial F_i}{\partial \lambda_j} \delta \lambda_j = - \sum_{I=1}^{N}\frac{\partial F_i}{\partial \mu_I} \delta \mu_I~,\quad i=1,\dots,5~.
\end{equation} Since the type \ref{item:general} solution is a fixed point solution of the minimal model, we may safely assume that the determinant of $\partial F_i/\partial\lambda_j$ is non-zero. This would give a solution $(\delta { \ell }, \delta \lambda, \delta { g_1 }, \delta v, \delta d)$ in terms of the $\delta\mu_I$.  Since the determinant is a continuous function on the space of $5\times5$ matrices, the determinant will continue to be non-zero for small enough finite deformations. The deformation can thus be integrated by parallel transport, although the solution might depend on the path of the deformation.
Additionally, how small the finite deformation needs to be depends on the specifics of the higher order terms. For instance, if we simply vary only $\mu$ and $M$ then the inequality \eqref{eq:inequality} must be kept in order to preserve the solution.

A similar analysis can be performed for the anti-de Sittter solution type \ref{item:AdS}: higher order terms will only correct the radius $\ell$ of the anti-de Sitter solution and its scalar $D=d$. This shows that the gauge fields will remain unexcited in the space of warped anti-de Sitter backgrounds for small enough corrections from anti-de Sitter space. A similar analysis can be performed on the purely gravitational solution type \ref{item:pureTMG} without exciting the gauge fields. The same analysis however cannot be performed for the type \ref{item:special} solution. This is because the equivalent linearized matrix \eqref{eq:linmat}  at $M=4\mu$ for this solution has zero determinant, because the solution is a line parametrized by $\lambda$. The ensuing analysis would depend crucially on the specific linearized theory parameters that are added and this is beyond the scope of this analysis.

One may now broadly comment on the naturalness of the timelike and spacelike warped anti-de Sitter solutions in more general theories. A theory with a metric, $N_V$ vectors and $N_S$ scalars will be parametrized by $2+N_V+N_S$ constants. This is the same number of isometry-invariant components of the field equations. Whether the non-linear system of algebraic equations has a solution depends on the details of the theory. Nevertheless, the existence of a solution can always be explained in this framework naturally. The same is not true if we add, for instance, a higher-spin tensor. For instance, a spin-two symmetric field $T_{\mu\nu}$ will be parametrized by two constants, but will typically obey two field equations, $\Box T = \ldots\,$ and $\nabla^{\nu}T_{\mu\nu}=0$, yielding ultimately three algebraic equations. The system would be in principle overdetermined.

\section{As Supersymmetric Backgrounds}
\label{sec:assusy}
In this section, we ask how much (2,0) supersymmetry a warped anti-de Sitter background admits. For the spacelike warped anti-de Sitter we find the spin coefficients (all in flat coordinates)
\begin{equation}
\omega_{\underline{1}\underline{2}\underline{0}} = \frac{\lambda}{{ \ell }} , \quad
\omega_{\underline{0}\underline{1}\underline{2}} = \frac{\lambda}{{ \ell }} ,\quad
\omega_{\underline{2}\underline{1}\underline{0}} = \frac{\lambda^2-2}{\lambda { \ell }}
\end{equation}
and calculate the curvature two-form
\begin{equation}\label{eq:curvature1}
R^{\nabla}_{\underline{a}\underline{b}} = \grad \omega_{\underline{a}\underline{b}} + \omega_{\underline{a}\underline{c}}\wedge \omega^{\underline{c}}{}_{\underline{b}} = \begin{cases}
-\frac{\lambda^2}{{ \ell }^2} \, \theta_{\underline{a}} \wedge \theta^{\underline{2}} 
&\text{if }\underline{a}=\underline{0},\underline{1} \text{ and } \underline{b}=\underline{2}\\
\frac{4-3\lambda^2}{{ \ell }^2} \,  \theta^{\underline{0}} \wedge \theta^{\underline{1}} &\text{if }[\underline{a}\underline{b}]=[\underline{0}\underline{1}]
\end{cases}~.
\end{equation}
The Killing spinor equation is
\begin{equation}
\begin{aligned}\label{eq:ks1}
\nabla_{\mu}\epsilon &= \left( i V_{\mu} + i \gamma_{\mu} G + D \gamma_{\mu} \right) \epsilon \\
&=  \left( i(\frac{v}{2}-{ g_1 })\gamma_{\mu}\theta^{\underline{2}} + i \frac{v}{2} \theta^{\underline{2}}\gamma_{\mu} + d \gamma_{\mu} \right) \epsilon~.
\end{aligned}\end{equation}
and we will use the relation\footnote{We clarify that $\nabla$ acts on the one-form $\theta^{\underline{2}}$ so that $\nabla_{\mu} \theta^{\underline{a}}=\omega_{\mu \underline{b}}{}^{\underline{a}}\theta^{\underline{\beta}}$ and $k=\theta^{\underline{2}}=\gamma^{\underline{2}}$ acts on spinors.} in the image of the Clifford algebra representation
\begin{equation}\label{eq:nablakprop1}
\nabla_\mu \theta^{\underline{2}} = \frac{\lambda}{{ \ell }} i_\mu ( \theta^{\underline{0}} \wedge \theta^{\underline{1}} ) = - \frac{\lambda\zeta}{2{ \ell }}[\gamma_{\mu},\theta^{\underline{2}}]~.
\end{equation}
We act on \eqref{eq:ks1} with $\nabla_{\nu}$
\begin{equation}\label{eq:nabla2eps}
\begin{aligned}
\nabla_{\nu}\nabla_{\mu}\epsilon &=   \left( i(\frac{v}{2}-{ g_1 })\gamma_{\mu} \left(\nabla_{\nu}\theta^{\underline{2}}\right) + i \frac{v}{2} \left(\nabla_{\nu}\theta^{\underline{2}}\right) \gamma_{\mu} \right) \epsilon \\
&+   \left( i(\frac{v}{2}-{ g_1 })\gamma_{\mu}\theta^{\underline{2}} + i \frac{v}{2} \theta^{\underline{2}}\gamma_{\mu} + d \gamma_{\mu} \right) \nabla_{\nu} \epsilon~,
\end{aligned}
\end{equation}
antisymmetrize in $[\mu,\nu]$ and use \eqref{eq:curvature1}, \eqref{eq:ks1} and \eqref{eq:nablakprop1}. In particular, the curvature acts in the spin algebra as $[\nabla_{\mu},\nabla_{\nu}]=1/4\, R_{\mu\nu ab}\gamma^{ab}$. This way, we arrive at the first-order integrability conditions
\begin{align}
\left[\left( \frac{\lambda^2}{2{ \ell }^2}-2d^2 \right) \gamma_{\underline{2}} + i { g_1 } \left( 2d
- \frac{\lambda\zeta}{{ \ell }} \right) \right]\epsilon &=0\\
\left[ \left( \frac{4-3\lambda^2}{2{ \ell }^2}-2\left(d^2+{ g_1 }^2\right) \right)\gamma_{\underline{2}} + 2 i \left(v-{ g_1 }\right) \frac{\lambda\zeta}{{ \ell }} \right] \epsilon &=0~.
\end{align}
The condition for maximal supersymmetry is either $v={ g_1 }=0$, $\lambda=1$ and  $|d|=1/(2{ \ell })$, i.e. anti-de Sitter, or $v\neq0$ and
\begin{align}
v &= { g_1 } \label{eq:maxsp1}\\
d&=\frac{\lambda\zeta}{2{ \ell }} \label{eq:maxsp2}\\
{ g_1 }^2&= \frac{1-\lambda^2}{{ \ell }^2}~. \label{eq:maxsp3}
\end{align}
Full supersymmetry requires a squashed $\lambda<1$ spacelike deformation. 
The conditions for half supersymmetry is derived from a spinor $\gamma_{\underline{2}}\epsilon=\pm\epsilon$ and are
\begin{align}
\label{eq:twofac1}
\left( \frac{\lambda\zeta}{2{ \ell }} - d \right) \left( -i{ g_1 } \pm \left( d+ \frac{\lambda\zeta}{2{ \ell }}  \right) \right)&=0\\
iv = i{ g_1 } \pm \frac{{ \ell }\zeta}{\lambda}\left( d^2+{ g_1 }^2 -\frac{4-3\lambda^2}{4{ \ell }^2}\right)&~.\label{eq:twofac2}
\end{align}
However, the half-supersymmetry conditions are void\footnote{If we were to allow imaginary $v$ and $g$ then higher-order integrability conditions require the second factor of \eqref{eq:twofac1} to vanish.} for real values of $v$ and ${ g_1 }$. A spacelike warped anti-de Sitter has either maximal supersymmetry or none. Regarding the minimal solutions, the spacelike warped type \ref{item:special} solution is indeed maximally supersymmetric.

For the timelike warped anti-de Sitter background we find the spin coefficients (again all in flat spin coefficients)
\begin{equation}
\omega_{\underline{1}\underline{2}\underline{0}} = \frac{\lambda}{{ \ell }} , \quad
\omega_{\underline{2}\underline{0}\underline{1}} = \frac{\lambda}{{ \ell }}, \quad
\omega_{\underline{0}\underline{1}\underline{2}} = \frac{2-\lambda^2}{\lambda { \ell }}
\end{equation}
and curvature
\begin{equation}
R^{\nabla}_{\underline{a}\underline{b}} = \grad \omega_{\underline{a}\underline{b}} + \omega_{\underline{a}\underline{c}}\wedge \omega^{\underline{c}}{}_{\underline{b}} =
\begin{cases}
\frac{\lambda^2}{{ \ell }^2}\theta^{\underline{0}} \wedge \theta^{\underline{b}} , &
\text{if }\underline{a}=\underline{0}\text{ and }\underline{b}=\underline{1},\underline{2}\\
\frac{3\lambda^2-4}{{ \ell }^2} \theta^{\underline{1}} \wedge \theta^{\underline{2}}
&\text{if }[\underline{a}\underline{b}]=[\underline{0}\underline{1}]
\end{cases}~.
\end{equation}
The Killing spinor equation is now
\begin{equation}\label{eq:ks2}
\begin{aligned}
\nabla_{\mu}\epsilon 
&=  \left( i(\frac{v}{2}+{ g_1 })\gamma_{\mu}\theta^{\underline{0}} + i \frac{v}{2} \theta^{\underline{0}}\gamma_{\mu} + d \gamma_{\mu} \right) \epsilon~ 
\end{aligned}\end{equation}
and we use the relation (again in the Clifford module)
\begin{equation}
\nabla_\mu \theta^{\underline{0}} = \frac{\lambda}{{ \ell }} i_\mu ( \theta^{\underline{2}} \wedge \theta^{\underline{1}} ) = \frac{\lambda\zeta}{2{ \ell }}[\gamma_{\mu},\theta_{\underline{0}}]~.\label{eq:nablatheta0}
\end{equation}
A similar calculation to the one described previously yields the integrability conditions
\begin{align}
\left[ i{ g_1 }(2d-\frac{\lambda\zeta}{{ \ell }}) + ( 2d^2- \frac{\lambda^2}{2{ \ell }^2})\gamma_{\underline{0}} \right]\epsilon &=0
\\
\left[ 
- i \frac{\lambda\zeta}{{ \ell }}(v+{ g_1 }) + \left( \frac{3\lambda^2-4}{4{ \ell }^2}-{ g_1 }^2+d^2 \right) \gamma_{\underline{0}}\right]\epsilon &=0
\end{align}
The maximal supersymmetry condition is either $v={ g_1 }=0$, $\lambda=1$ and  $|d|=1/(2{ \ell })$, i.e. anti-de Sitter, or $v\neq0$ and
\begin{align}
v &= -{ g_1 } \\
d&=\frac{\lambda\zeta}{2{ \ell }} \\
{ g_1 }^2&= \frac{\lambda^2-1}{{ \ell }^2}~.
\end{align}
We deduce that the special type \ref{item:special} timelike warped anti-de Sitter minimal solution satisfies the conditions for maximal supersymmetry.

The half-supersymmetry conditions are derived from the projection  $i\gamma_{\underline{0}}\epsilon=\pm\epsilon$ and are
\begin{align}
\left( \frac{\lambda\zeta}{2{ \ell }} - d \right) \left(-{ g_1 } \pm  \left( \frac{\lambda\zeta}{2{ \ell }}  + d \right)\right) &=0
\label{eq:twofac3}\\
v+{ g_1 } \pm \frac{{ \ell }\zeta}{\lambda}\left( \frac{3\lambda^2-4}{4{ \ell }^2}+d^2-{ g_1 }^2 \right)&=0~.\label{eq:twofac4}
\end{align}
However, the first-order integrability condition of the projection
\begin{equation}
i \theta_{\underline{0}} \epsilon = \pm \epsilon~,
\end{equation}
for instance acting on this equation with $\nabla_{\underline{i}}$, $\underline{i}=\underline{1},\underline{2}$, and using \eqref{eq:ks2} and \eqref{eq:nablatheta0} again, yields
\begin{align}
\label{eq:cond1}
g_1 &= \pm \left( d + \frac{\lambda\zeta}{2\ell}\right)~,\\\intertext{i.e. only the second factor of \eqref{eq:twofac3} needs to vanish, in which case \eqref{eq:twofac4} becomes}
v &= \mp \frac{\lambda^2-1}{2\lambda}\zeta~.
\label{eq:cond2}
\end{align}
The necessary and sufficient conditions for half-supersymmetry are \eqref{eq:cond1} and \eqref{eq:cond2}. The general type \ref{item:general} solution only satisfies the condition for half supersymmetry with
\begin{equation}
g_1 = \pm \frac{3M+4\mu}{16}~.
\end{equation}
That is, both signs of $g_1$ are half-supersymmetric.

The analysis of the null warped backgrounds is given in the appendix \ref{sec:null}. In general, the background is supersymmetric only if 
\begin{equation}
d = - \frac{\zeta}{2\ell} \quad\text{and}\quad v\ell=-2 g_1\zeta~.
\end{equation}
It is generically half-supersymmetric, unless the deformation has negative deformation and 
\begin{equation}
g_1 = \frac{1}{2}~.
\end{equation}
This null warped anti-de Sitter is maximally supersymmetric as the limit of the maximally supersymmetric warped spacelike and timelike anti-de Sitter, a procedure that may only enhance the kernel of the supercovariant derivative $D_{\mu}$~\cite{geroch_limits_1969,Blau:2002mw}. 

In summary, only the warped anti-de Sitter solutions of type \ref{item:special} are maximally supersymmetric, but these are only solutions for the minimal model with $\mu=M/4$. This is still in agreement with \cite{knodel_rigid_2015}. In their work, they find all backgrounds that are maximally supersymmetric, but the equations of a theory are not used. That is, warped anti-de Sitter is found to be among the  maximally supersymmetric backgrounds but only for certain values of the size of $\lambda$, $\ell$ etc, which in turn might not be solutions to a given theory. Other values might be solutions that allow less or none supersymmetry. Both caveats appear in the minimal model, as also noted in \cite{prep}. If we identify backgrounds up to rescalings of $\ell$ then we also see that there are essentially no other supersymmetric backgrounds than the minimal model solutions. That is, up to rescalings of $\ell$ the right complement in figure \ref{fig:venn} is actually empty, for instance half-supersymmetric solutions are parametrized by two parameters, $\mu$ and $M$ or $\lambda$ and $d$. Of course, non-supersymmetric minimal model solutions and non-supesymmetric backgrounds that are not solutions always exist.

\section{As Supersymmetric Quotients}
\label{sec:asquotients}
The regular isometric quotients of spacelike, timelike and null warped anti-de Sitter were investigated in \cite{anninos_warped_2009}. The identification under an isometry, 
\begin{equation}
e^{2\pi \partial_{\theta}}p \approx p \quad \text{for all }p\in M \quad\text{where}\quad\partial_{\theta}\in\mathfrak{iso}_k~,
\end{equation}
is required to either define a quotient without pathologies, or if such pathologies exist, namely closed causal curves, that they are hidden behind an absolute horizon as in the BTZ construction of \cite{Banados:1992gq}. One may then ask how much $(2,0)$ supersymmetry the regular quotients will preserve. For the supersymmetric BTZ quotients see \cite{izquierdo_supersymmetric_1995}.

Our treatment will not be exhaustive. We will concentrate on two quotients:
\begin{enumerate}
\item The so-called self-dual quotient \cite{jugeau_accelerating_2011}, see also \cite{coussaert_selfdual_1994}, with $\partial_{\theta} =T\, L_2$ on the maximally supersymmetric spacelike warped anti-de Sitter.
\item The ``asymmetric vacuum'' quotient, similar to the vacuum used in \cite{compere_boundary_2009}, which corresponds to the quotient of half-maximal timelike squashed $\lambda<1$ anti-de Sitter with $\partial_{\theta} = R_0+R_2$. The minimal model solution for $\lambda<1$ requires 
\begin{equation}
\mu(2\mu-3M)<0~.
\end{equation}
\end{enumerate}
In both cases there are no pathologies~\cite{anninos_warped_2009}. However, the quotients might break some or all supersymmetry. In \cite{prep} a specific Killing spinor is explicitly solved for and shown to be independent of the angle $\theta$ in both of the above quotients. Our purpose is to see precisely \textit{how many} Killing spinors survive and highlight some subtleties in the problem.

We will now see that the first quotient, the self-dual quotient, is in fact maximally supersymmetric. This is because $\partial_{\theta}$ is central in the Killing superalgebra and in particular it acts trivially on the space of Killing spinors $\CC^2$. Indeed, the spinorial Lie derivative along $k=2/(\lambda\ell)L_2=\theta^{\underline{2}}$ on any Killing spinor $\epsilon$ is 
\begin{equation}\label{eq:liekeps1}
\begin{aligned}
\lie_{k} \epsilon &= \left(\nabla_{k} + \frac{1}{4} \left( \grad k \right) \right)\epsilon \\
&=  \left( i(\frac{v}{2}-{ g_1 })\theta^{\underline{2}}\theta^{\underline{2}} + i \frac{v}{2} \theta^{\underline{2}} \theta^{\underline{2}} + d\, \theta^{\underline{2}} + \frac{1}{4} \frac{2 \lambda}{\ell} \theta^{\underline{0}} \wedge \theta^{\underline{1}} \right) \epsilon \\
&=  \left( i\left(v-{ g_1 }\right) + \left( d - \frac{ \lambda\zeta}{2\ell} \right)\theta^{\underline{2}}\right) \epsilon
\\
&=0~,
\end{aligned}
\end{equation} 
where we use the definition of a Killing spinor \eqref{eq:ks1} to replace $\nabla_k \epsilon$, the maximal supersymmetry conditions \eqref{eq:maxsp1}-\eqref{eq:maxsp3}, and the identity $\theta^{\underline{0}}\wedge\theta^{\underline{1}}=-\theta^{\underline{2}}$ from  \eqref{eq:hodgegamma}. As a result, for $\partial_{\theta}=T L_2$ we have that $\lie_{\partial_{\theta}}\epsilon=0$ and all Killing spinors survive the self-dual quotient. In fact, all of its Killing superalgebra, i.e. including its Killing vectors, commute with $\partial_{\theta}$ and hence are preserved by the quotient. Let us comment that the Killing superalgebra is not entirely trivial. The Killing spinors transform under $\mathfrak{sl}(2,\RR)_L$ and the odd-odd bracket of two Killing spinors should square into $\mathfrak{sl}(2,\RR)_L$.  Note that $\partial_{\theta}$ leaves all of the background fields invariant, including the $U(1)_R$ gauge field $V$. 

In order to discuss the supersymmetry preserved by the second quotient, that on the half-supersymmetric space by $R_0+R_1$, one should calculate the action of $R_{0}+R_1 \in \mathfrak{iso}_k$ on its complex one-dimensional space of Killing spinors $\CC$. 
 In order to do this, we will need the relation
\begin{equation}\label{eq:dR}
\grad R_a = - \frac{2\lambda}{\ell}\zeta \ast R_a - 8 \frac{\lambda^2-1}{\ell^2\lambda^2}\zeta \,\theta^{\underline{0}}(R_a) \, \ast L_0~,
\end{equation}
for the derivatives of the left-action Killing vectors $R_a$ in the timelike warped anti-de Sitter background. A sketch of the proof of \eqref{eq:dR} is given in appendix \ref{app:dR}. Note here that such a form is expected from representation theory and only the constants in \eqref{eq:dR} have to be calculated. Then, by using the half-supersymmetry conditions of timelike warped anti-de Sitter, \eqref{eq:cond1} and \eqref{eq:cond2}, the action turns out to be trivial:
\begin{equation}
  \begin{aligned}
\lie_{R_a} \epsilon & =  \left(\nabla_{R_a} + \frac{1}{4} \left( \grad R_a \right) \right)\epsilon \\
&=  \left( i(\frac{v}{2}+{ g_1 })R_a\theta^{\underline{0}} + i \frac{v}{2} \theta^{\underline{0}} R_a + d\, R_a +  \frac{ \lambda\zeta}{2\ell} R_a + \frac{\lambda^2-1}{\lambda\ell}\zeta \theta^{{\underline{0}}}(R_{a}) \right) \epsilon \\
&=  0~.
  \end{aligned}
\end{equation}
Again, the quotient preserves its half supersymmetry. At the same time, the isometry algebra of the quotient is broken to the two commuting elements $R_0+{R_1}$ and $L_2$. Note that the quotient vector $\partial_{\theta}=R_0+R_1$ preserves all fields including the gauge field $V$.

This concludes the study of the two quotients. In a sense we have been lucky, since the two quotients discussed preserve all the fields and the Killing spinors. This was not expected a priori and a similar calculation to \eqref{eq:liekeps1} shows that in the case of the timelike half-supersymmetric warped anti-de Sitter background, the Killing vector $k=2/(\ell\lambda)L_0$ does act on its Killing spinor
\begin{equation}\label{eq:lastk}
\lie_k \epsilon  = \mp i \frac{\zeta}{\lambda\ell}\epsilon~.
\end{equation}
In fact, a different $U(1)_R$ gauge choice\footnote{This is precisely what is happening in the formalism of \cite{prep}.}, under which the Killing spinor transforms as  \eqref{eq:KStransU1}, will render the Killing spinor invariant under $k$. That is, \eqref{eq:lastk} is replaced with $\lie_k\epsilon=0$ in a different gauge. However, in this new gauge choice the Killing vector $k$ would not leave invariant the gauge field $V$, as would follow from consistency of \eqref{eq:Vtrans}, \eqref{eq:lastk} and \eqref{eq:KStransU1}. Although this did not happen in the two previous examples, more generally and if the fundamental group is not trivial then a supersymmetric quotient may be reached by using a $U(1)_R$ patching under which both $V$ and the Killing spinors transform. Alternatively, the quotient might require a non-trivial spin structure, which may even include Lorentz rotations. The identification under $e^{2\pi L_0}$ in the timelike half-maximal warped anti-de Sitter discussed here is in fact supersymmetric and yields an anti-periodic spin structure $\epsilon\mapsto-\epsilon$. In any case, this discussion is not necessary since $k=2/(\ell\lambda) L_0$ is timelike and the quotient has closed timelike curves.


\section{Conclusion}
In this note we focused on warped anti-de Sitter backgrounds in $(2,0)$ supergravity. We gave conditions of when they preserve half supersymmetry and we also explained why one should indeed expect to find such solutions in the minimal model and its deformations. Similar statements can be made for $(1,1)$ supergravity and indeed warped anti-de Sitter features in the supersymmetric solution of \cite{deger_supersymmetric_2013}. A simple counting shows that these solutions are in a sense natural. We hope to have given here a useful and clarifying answer about the ubiquity of these geometries, as was our motivation.

All maximally supersymmetric backgrounds of $(2,0)$ supergravity where derived in \cite{knodel_rigid_2015}. Besides anti-de Sitter and Minkowski, the rest also have the form of a Hopf-like\footnote{That is, the field-strength of the fibration is proportional to the volume form on the base space and the dilaton of the fibration is constant.} fibration over a two-dimensional maximally symmetric space of either signature
\begin{align}
S^2 \tilde{\times} \RR &\quad \text{lorentzian sphere}\\
\RR^2 \tilde{\times} \RR& \quad \text{warped flat}\\
H^2 \tilde{\times} \RR &\quad \text{timelike warped anti-de Sitter}\\
\AdS_2 \tilde{\times} \RR &\quad \text{spacelike warped anti-de Sitter}~,
\end{align}
or else they are a generalization of a pp-wave. The lorentzian sphere is, in a similar sense to our definition, a vector-like deformation\footnote{In euclidean signature this is a biaxial squashing/stretching of the sphere. In the Bianchi three-dimensional group classification, two of the diagonal parameters on the $\mathrm{SU(2)}$ metric are equal as in \eqref{eq:wads}.} of the round sphere. 
Very similar results to what were produced here are expected for the lorentzian sphere. Indeed, in \cite{imamura_n2_2012} the form of the superderivative on the squashed sphere that admits maximal supersymmetry is very similar to \eqref{eq:ks1} and \eqref{eq:ks2}. For instance, one may ask whether the maximal supersymmetric lorentzian sphere is a solution to the minimal or some other model, because the method of \cite{prep} did not produce the maximally supersymmetric lorentzian sphere. 

We have not been exhaustive in all of our treatment in this note. For instance, there are still some healthy quotients of timelike squashed anti-de Sitter without horizons and some other quotients of the null warped anti-de Sitter, see \cite{anninos_warped_2009}. Another interesting question that arises from our exposition is whether it is natural to expect only supersymmetric minimal solutions when the gauge fields are turned on. For instance, if one deforms the theory as in the end of section \S\ref{sec:assolutions}, one expects that the supersymmetry conditions of section \S\ref{sec:assusy} will not hold anymore. So why are all non-gravitational warped anti-de Sitter solutions of the minimal model supersymmetric to begin with? It is also not clear here under what conditions a supersymmetric theory deformation preserves the supersymmetry of the deformed solutions.

\subsection*{Acknowledgments}
The author acknowledges support from the Scientific and Technological Research Council of
Turkey (T\"UB\.ITAK) project 113F034.

\appendix
\section{Null Warped anti-de Sitter}
\label{sec:null}
We repeat here the analyses for the case of null warped anti-de Sitter. Recall that we differentiate two backgrounds with a minus or plus sign, see \eqref{eq:metricnulltwo}. The Ricci tensor and Cotton tensor decompose into
\begin{align}
\text{Ricci} &= -\frac{2}{{ \ell }^2} g \mp \frac{4}{{ \ell }^2} k \otimes k
\\
\text{Cotton} &= \mp \frac{12}{{ \ell }^3}\zeta  k \otimes k~.
\end{align}
The minimal model equation of motion for $G$ gives
\begin{equation}
(\mu+2d)g_1= -\frac{2}{{ \ell }}\zeta g_1~,
\end{equation}
the equation of motion for $V$ solves for $v$,  
while the Einstein equation of motion decomposes into
\begin{align}
4 d^2{ \ell^2 } &=1\\
\mp \frac{4}{{ \ell }^2} \mp \frac{12 \zeta}{\mu { \ell }^3} + 32  \frac{{ g_1 }^2}{{ \ell }^2} &=0
\end{align}
and $d=M/8$ as before. With the radius $\ell$ fixed, the only solutions are
\begin{enumerate}
\item If $\mu\ell=-3\zeta$, then both positive and negative purely gravitational ${ g_1 }=0$  null warped anti-de Sitter are allowed.
\item  If $M=4\mu$, then the negative deformation is allowed with $2d\ell=-\zeta$ and $|g_1|=1/2$. 
\end{enumerate}
Note that only the last solution has the vectors turned on, and both solutions impose a restriction on $\mu$ or $M$, $4|\mu|=3|M|$ and $M=4\mu$ respectively. 

In order to write the Killing spinor equation, we define the frame
\begin{equation}
\theta^{{a}} = \frac{{ \ell }}{2} \times \begin{cases} \tau_2 & {a} = {2}
\\
\frac{1}{\sqrt{2}} \left( \tau_0 + \tau_1 \right)
&{a}=+
\\
 \frac{1}{\sqrt{2}} \left( -\tau_0 + \tau_1 \pm \left( \tau_0 +\tau_1\right)\right)
&{a}=- \end{cases}~,
\end{equation}
use the metric with $\eta_{+-}=1$ and $\gamma^{+-2}=-\zeta$. We find the spin coefficients (in flat indices)
\begin{align}
\omega_{{2}-} &= - \frac{1}{\ell}\theta^+ \\
\omega_{+-} &= \frac{1}{\ell} \theta^{\underline{2}}\\
\omega_{2+} &= \frac{1}{\ell} \theta^- \mp \frac{4}{\ell} \theta^+~.
\end{align}
The curvature two-form is
\begin{align}
R_{+-} &= - \frac{1}{\ell^2} \theta^- \wedge \theta^+ \\
R_{-2} &= - \frac{1}{\ell^2} \theta^+ \wedge \theta^2 \\
R_{+2} &= - \frac{1}{\ell^2} \theta^- \wedge \theta^2 \mp \frac{8}{\ell^2} \theta^+ \wedge \theta^2~. 
\end{align}
After some calculation, the integrability conditions, respectively for $[D_{2},D_+]$, $[D_+,D_-]$ and $[D_2,D_-]$, are
\begin{align}
\Big(\frac{1}{2\ell^2}\gamma^{-2} \pm \frac{4}{\ell^2} \gamma^{+2}\Big)\epsilon
&= \Big(  \frac{i}{\ell} \left(a-b\right)\gamma^{-+} + 2 \frac{i}{\ell} \left(a+b\right) \\ \notag &{}\quad - 2 \zeta (a-b)^2 \gamma^+ - 2 d i (a-b) \gamma^2 - 2 \zeta d^2 \gamma^- \Big)\epsilon\\
-\frac{1}{2\ell^2} \gamma^{-+}\epsilon &= 
\left( -\zeta\frac{i}{\ell}\left(a-b\right) \gamma^+ + 2i d \gamma^+ \left(a-b\right)- 2 d^2 \gamma^2 \zeta\right)\epsilon\\
- \frac{1}{2\ell^2}\gamma^{2+}\epsilon &= 2 d^2 \zeta \gamma^+ \epsilon~,
\end{align}
where
\begin{align}
a &= \frac{\sqrt{2}}{2}v + \frac{2\sqrt{2}g_1\zeta}{\ell} \\
b &= \frac{\sqrt{2}v}{2}
\end{align}
and we have also used repeatedly relations such as $\gamma^{2+}=-\zeta\gamma^{+}$ and $\gamma^{2-}=\zeta\gamma^{-}$. Half supersymmetric solutions arise from the projection $\gamma_-\epsilon=0$ and require $a=-b$ and $d=\frac{\zeta}{2\ell}$. If furthermore $a^2=1/(2\ell^2)$ for the negative sign null deformation, the supersymmetry is enhanced to maximal. 

\section{Derivatives of $R_a$}
\label{app:dR}
We will only sketch the proof of \eqref{eq:dR}. 
By using the formula for a one-form $\omega_{[1]}$
\begin{equation}\label{eq:domega}
\grad \omega_{[1]} (X,Y) = \lie_X \left(\omega_{[1]}(Y)\right) - \lie_Y \left(\omega_{[1]}(X)\right) - \omega_{[1]}([X,Y])~,
\end{equation}
we calculate
\begin{equation}
\grad (R_0+R_1)(R_0+R_1,L_2) = 0~,
\end{equation}
and by using 
\begin{equation}\label{eq:ixy}
i_{X\wedge Y} \omega_{[2]} = \left. X \wedge Y \wedge \omega_{[2]} \right|_{\dvol}
\end{equation}
for a two-form such as $\omega_{[2]} = \grad (R_0+R_1)$, we are able to show that
\begin{equation}\label{eq:dRabform}
\grad (R_0+R_1) = a\, \ast (R_0 + R_1 ) + b \, \ast L_2~.
\end{equation}
This is indeed of the form in \eqref{eq:dR} and it remains to find $a$ and $b$. Contracting $\grad(R_0+R_1)$ again by using \eqref{eq:domega}, e.g. with
\begin{align}\label{eq:dRxy1}
\grad (R_0 + R_1) ( R_0 +R_1, R_2) &= - 2 g(R_0+R_1, R_0+ R_1)\\
\grad  (R_0 + R_1) (L_2, R_2) &= g(R_{0}+R_1,L_2)~,
\label{eq:dRxy2}
\end{align}
we may calculate the left-hand side of the above equations in terms of $\theta^{\underline{0}}(R_0+R_1)$ and $\theta^{\underline{0}}(R_2)$ with the use of
\begin{equation}
g = g_{\AdS} - \frac{\lambda^2-1}{\lambda^2} \left( \theta^{\underline{0}} \right)^2~,
\end{equation}
because the $R_a$ are orthonormal for the metric $g_{\AdS}$. 
It remains to prove that 
\begin{equation}\label{eq:dvol2}
\left( R_0 + R_1 \right) \wedge R_2 \wedge L_0 = \zeta \frac{\ell^2}{4} \theta^{\underline{0}}(R_0+R_1)
\end{equation}
in order to combine \eqref{eq:dRxy1} and \eqref{eq:dRxy2} with \eqref{eq:dRabform} and \eqref{eq:dvol2} so that one may solve for the constants $a$ and $b$. The result is then given by \eqref{eq:dR}. Similarly, for the spacelike warped metric, one may show that
\begin{equation}
\grad R_a = - \frac{2\lambda \zeta}{\ell} \ast R_a + 8 \frac{\lambda^2-1}{\ell^2\lambda^2} \theta^{\underline{2}}(R_a) \ast L_2~.
\end{equation}
By using the equation above, one may show how the $R_a$ acts on spacelike warped anti-de Sitter Killing spinors trivially. 

\providecommand{\href}[2]{#2}\begingroup\raggedright\endgroup
\end{document}